\documentclass{article}
\usepackage{fortschritte}
\usepackage{amssymb}
\usepackage{amsmath}
\usepackage{graphicx}
\def\beq{\begin{equation}}                     %
\def\eeq{\end{equation}}                       %
\def\bea{\begin{eqnarray}}                     
\def\eea{\end{eqnarray}}                       
\def\Ket{\rangle}

\newcommand{\ket}[1]{\ensuremath{\left| #1 \right\rangle}}

\newcommand{\eval}[1]{\ensuremath{\left\langle #1 \right\rangle}}

\newcommand{\perm}[1]{
  {P_{\!\scriptscriptstyle #1}}
}

\newcommand{\fourpoints}{  
  \put(0,0){\circle*{1.5}}\put(5,0){\circle*{1.5}}\put(10,0){\circle*{1.5}}\put(15,0){\circle*{1.5}} 
}

\newcommand{\diagAone}{
\begin{picture}(19,7)
  \put(0,1.5){
  \begin{picture}(15,10)
  \fourpoints
  \qbezier[100](0,0)(2.5,0)(5,0)
  \end{picture}}
\end{picture}
}

\newcommand{\diagAtwo}{
\begin{picture}(19,7)
  \put(0,1.5){
  \begin{picture}(15,10)
  \fourpoints
  \qbezier[100](0,0)(2.5,0)(5,0)
  \qbezier[100](10,0)(12.5,0)(15,0)
  \end{picture}}
\end{picture}
}

\newcommand{\diagBtwo}{
\begin{picture}(19,7)
  \put(0,1.5){
  \begin{picture}(15,10)
  \fourpoints
  \qbezier[100](0,0)(2.5,0)(5,0)
  \qbezier[100](5,0)(7.5,0)(10,0)
  \end{picture}}
\end{picture}
}

\newcommand{\diagCtwo}{
\begin{picture}(19,7)
  \put(0,1.5){
  \begin{picture}(15,10)
  \fourpoints
  \qbezier[100](0,0)(2.5,4)(5,0)
  \qbezier[100](0,0)(2.5,-4)(5,0)
  \end{picture}}
\end{picture}
}

\newcommand{\diagAthree}{
\begin{picture}(19,7)
  \put(0,1.5){
  \begin{picture}(15,10)
  \fourpoints
  \qbezier[100](0,0)(2.5,4)(5,0)
  \qbezier[100](0,0)(2.5,-4)(5,0)
  \qbezier[100](5,0)(7.5,0)(10,0)
  \end{picture}}
\end{picture}
}

\newcommand{\diagBthree}{
\begin{picture}(19,7)
  \put(0,1.5){
  \begin{picture}(15,10)
  \fourpoints
  \qbezier[100](0,0)(2.5,0)(5,0)
  \qbezier[100](5,0)(7.5,0)(10,0)
  \qbezier[100](5,0)(10,7)(15,0)
  \end{picture}}
\end{picture}
}

\newcommand{\diagCthree}{
\begin{picture}(19,7)
  \put(0,1.5){
  \begin{picture}(15,10)
  \fourpoints
  \qbezier[100](0,0)(2.5,0)(5,0)
  \qbezier[100](5,0)(7.5,0)(10,0)
  \qbezier[100](10,0)(12.5,0)(15,0)
  \end{picture}}
\end{picture}
}

\newcommand{\diagDthree}{
\begin{picture}(19,7)
  \put(0,1.5){
  \begin{picture}(15,10)
  \fourpoints
  \qbezier[100](0,0)(2.5,4)(5,0)
  \qbezier[100](0,0)(2.5,-4)(5,0)
  \qbezier[100](0,0)(2.5,0)(5,0)
  \end{picture}}
\end{picture}
}

\newcommand{\diagEthree}{
\begin{picture}(19,7)
  \put(0,1.5){
  \begin{picture}(15,10)
  \fourpoints
  \qbezier[100](0,0)(2.5,4)(5,0)
  \qbezier[100](0,0)(2.5,-4)(5,0)
  \qbezier[100](10,0)(12.5,0)(15,0)
  \end{picture}}
\end{picture}
}

\newcommand{\diagFthree}{
\begin{picture}(19,7)
  \put(0,1.5){
  \begin{picture}(15,10)
  \fourpoints
  \qbezier[100](0,0)(2.5,0)(5,0)
  \qbezier[100](5,0)(7.5,0)(10,0)
  \qbezier[100](0,0)(5,7)(10,0)
  \end{picture}}
\end{picture}
}

\newcommand{\diagAfour}{
\begin{picture}(19,7)
  \put(0,1.5){
  \begin{picture}(15,10)
  \fourpoints
  \qbezier[100](5,0)(7.5,0)(10,0)
  \qbezier[100](0,0)(5,7)(10,0)
  \qbezier[100](0,0)(5,-7)(10,0)  
  \qbezier[100](10,0)(12.5,0)(15,0)
  \end{picture}}
\end{picture}
}

\newcommand{\diagBfour}{
\begin{picture}(19,7)
  \put(0,1.5){
  \begin{picture}(15,5)
  \fourpoints
  \qbezier[100](0,0)(2.5,0)(5,0)
  \qbezier[100](10,0)(12.5,0)(15,0)
  \qbezier[100](5,0)(7.5,4)(10,0)
  \qbezier[100](5,0)(7.5,-4)(10,0)
  \end{picture}}
\end{picture}
}

\newcommand{\diagCfour}{
\begin{picture}(19,7)
  \put(0,1.5){
  \begin{picture}(15,10)
  \fourpoints
  \qbezier[100](0,0)(2.5,0)(5,0)
  \qbezier[100](5,0)(7.5,0)(10,0)
  \qbezier[100](0,0)(5,7)(10,0)
  \qbezier[100](0,0)(5,-7)(10,0)
  \end{picture}}
\end{picture}
}

\newcommand{\diagDfour}{
\begin{picture}(19,7)
  \put(0,1.5){
  \begin{picture}(15,10)
  \fourpoints
  \qbezier[100](0,0)(2.5,4)(5,0)
  \qbezier[100](0,0)(2.5,-4)(5,0)
  \qbezier[100](5,0)(7.5,0)(10,0)
  \qbezier[100](10,0)(12.5,0)(15,0)
  \end{picture}}
\end{picture}
}

\newcommand{\diagEfour}{
\begin{picture}(19,7)
  \put(0,1.5){
  \begin{picture}(15,10)
  \fourpoints
  \qbezier[100](0,0)(2.5,4)(5,0)
  \qbezier[100](0,0)(2.5,-4)(5,0)
  \qbezier[100](0,0)(2.5,7)(5,0)
  \qbezier[100](0,0)(2.5,-7)(5,0)
  \end{picture}}
\end{picture}
}

\newcommand{\diagFfour}{
\begin{picture}(19,7)
  \put(0,1.5){
  \begin{picture}(15,10)
  \fourpoints
  \qbezier[100](0,0)(2.5,4)(5,0)
  \qbezier[100](0,0)(2.5,-4)(5,0)
  \qbezier[100](5,0)(7.5,4)(10,0)
  \qbezier[100](5,0)(7.5,-4)(10,0)
  \end{picture}}
\end{picture}
}

\newcommand{\diagGfour}{
\begin{picture}(19,7)
  \put(0,1.5){
  \begin{picture}(15,10)
  \fourpoints
  \qbezier[100](0,0)(2.5,4)(5,0)
  \qbezier[100](0,0)(2.5,-4)(5,0)
  \qbezier[100](0,0)(2.5,0)(5,0)
  \qbezier[100](5,0)(7.5,0)(10,0)
  \end{picture}}
\end{picture}
}

\newcommand{\diagHfour}{
\begin{picture}(19,7)
  \put(0,1.5){
  \begin{picture}(15,10)
  \fourpoints
  \qbezier[100](0,0)(2.5,4)(5,0)
  \qbezier[100](0,0)(2.5,-4)(5,0)
  \qbezier[100](0,0)(2.5,0)(5,0)
  \qbezier[100](10,0)(12.5,0)(15,0)
  \end{picture}}
\end{picture}
}

\newcommand{\diagIfour}{
\begin{picture}(19,7)
  \put(0,1.5){
  \begin{picture}(15,10)
  \fourpoints
  \qbezier[100](0,0)(2.5,4)(5,0)
  \qbezier[100](0,0)(2.5,-4)(5,0)
  \qbezier[100](10,0)(12.5,4)(15,0)
  \qbezier[100](10,0)(12.5,-4)(15,0)
  \end{picture}}
\end{picture}
}

\newcommand{\diagJfour}{
\begin{picture}(19,7)
  \put(0,1.5){
  \begin{picture}(15,10)
  \fourpoints
  \qbezier[100](0,0)(2.5,0)(5,0)
  \qbezier[100](5,0)(7.5,0)(10,0)
  \qbezier[100](10,0)(12.5,0)(15,0)
  \qbezier[100](0,0)(5,7)(10,0)
  \end{picture}}
\end{picture}
}

\newcommand{\diagKfour}{
\begin{picture}(19,7)
  \put(0,1.5){
  \begin{picture}(15,10)
  \fourpoints
  \qbezier[100](0,0)(2.5,0)(5,0)
  \qbezier[100](5,0)(7.5,0)(10,0)
  \qbezier[100](10,0)(12.5,0)(15,0)
  \qbezier[100](0,0)(7.5,7)(15,0)
  \end{picture}}
\end{picture}
}

\begin {document}                 
\def\titleline{
%
%
%
A Note on Four-Point Functions in\\           
Logarithmic Conformal Field Theory            
}
\def\email_speaker{
{\tt 
%
%
flohr@itp.uni-hannover.de                    
}}
\def\authors{
%
%
%
%
%
Michael Flohr,\1ad\2ad$\!\!\!\!$\sp          
\phantom{mmmmmmmmmm}                         
Marco Krohn\2ad                              
\phantom{m}                                  
}
\def\addresses{
%
%
%
%
                                            %
\begin{tabular}{cc}                         %
\1ad Physikalisches Institut &              %
\2ad Institute for Theoretical Physics\\    %
University of Bonn &                        %
University of Hannover\\                    %
Nussallee 12, D-53115 Bonn, Germany &       %
Appelstr. 2, D-30167 Hannover, Germany      %
\end{tabular}                               
                                            %
                                            %
}
\def\abstracttext{
%
%
The generic structure of 4-point functions  
of fields residing in indecomposable        
representations of arbitrary rank is given. 
The presented algorithm is illustrated with 
some non-trivial examples and permutation   %
symmetries are exploited to reduce the      %
number of free structure-functions, which   %
cannot be fixed by global conformal         %
invariance alone.                           %
}
\large
\makefront
\section{Introduction}
During the last few years, logarithmic conformal field
theory (LCFT) has been established as a well-defined variety of
conformal field theories in two dimensions. The concept was considered in
its own right first by Gurarie \cite{Gurarie:1993xq}, Since then, a large
amount of work has appeared, see the reviews \cite{Flohr:2001zs,
Gaberdiel:2001tr} and references therein. The defining feature of a LCFT
is the occurrence of indecomposable representations which, in turn, may lead to
logarithmically diverging correlation functions. Thus, in the standard example
of a LCFT a primary field $\phi(z)$ of conformal weight $h$ has a so-called 
logarithmic partner field $\psi$ with the characteristic properties
\begin{equation}\label{eq:1}
        \eval{\phi(z)\phi(0)} = 0\,,\ \ \ \
        \eval{\phi(z)\psi(0)} = Az^{-2h}\,,\ \ \ \
        \eval{\psi(z)\psi(0)} =z^{-2h}\left(B - 2A\log(z)\right)\,.
\end{equation}
To this corresponds the fact that the highest weight state $\ket{h}$ 
associated to the primary field $\phi$ is the ground state of an irreducible 
representation which, however, is part of a larger, indecomposable, 
representation created from $|\tilde h\Ket$, the state associated to $\psi$. 
The conformal weight is the eigenvalue
under the action of $L_0$, the zero mode of the Virasoro algebra, which in
such LCFTs cannot be diagonalized. Instead, we have
\begin{equation}\label{eq:2}
        L_0\ket{h} = h\ket{h}\,,\ \ \ \ L_0|\tilde h\Ket = h|\tilde h\Ket
        + \ket{h}\,.
\end{equation}
Thus, the two states $\ket{h}$ and $|\tilde h\Ket$ span a Jordan cell of rank 
two with respect to $L_0$.
As can be guessed from eq.~(\ref{eq:1}), there must exist a zero mode which
is responsible for the vanishing of the 2-pt function of the primary field.
Another characteristic fact in LCFT is the existence of at least one field,
which is a perfect primary field, but whose operator product expansion (OPE)
with itself produces a logarithmic field. Such fields $\mu$ are called
pre-logarithmic fields \cite{Kogan:1997fd}. This is important, since in
many cases, the pre-logarithmic fields arise naturally forcing us then to
include the logarithmic fiels as well into the operator algebra. Note that
this implies that the fusion product of two irreducible representations is
not necessarily completely reducible into irreducible representations.
In fact, we know today quite a few LCFTs, where precisely this is the case,
such as ghost systems \cite{Krohn:2002gh}, WZW models at level zero or at
fractional level such as $\widehat{SU(2)}_{-4/3}$ \cite{Gaberdiel:2001ny,
Kogan:2001nj}, WZW models of 
supergroups such as $GL(1,1)$ \cite{Rozansky:1992td} or certain supersymmetric 
$c=0$ theories such as $OSP(2n|2n)$ or $CP(n|n)$ \cite{Gurarie:1999yx,
Read:2001pz}. Finally, many LCFTs are 
generated from free anticommuting fields such as the symplectic fermions
\cite{Kausch:2000fu}. LCFT enjoys numerous applications in condensed matter
physics, but it is important in string theory as well, e.g.~for the 
understanding of decaying $D$-branes \cite{Lambert:2003zr}.

In these notes, we generalize LCFT to the case of Jordan cells of arbitrary
rank, but we will restrict ourselves to the Virasoro algebra as the chiral
symmetry algebra to keep things simple. As has been shown in 
\cite{Flohr:2001tj,Flohr:2003tc}, the generic form of 1-, 2- and 3-pt functions
can be fixed up to structure constants under mild assumption on the structure of
the indecomposable representations. From this, the general 
structure of the OPE can then easily be obtained. However, in order to be
able to compute arbitrary correlation functions in LCFT, one at least needs
the 4-point functions such that crossing symmetry can be exploited.
Unfortunately, this turns out to be 
more complicated \cite{Flohr:1998ew,Flohr:2000mc}. In the following, we present
an algorithm with which the generic form of 4-point functions can be fixed
up to functions, which only depend on the globally conformal invariant
crossing ratio. In contrast to ordinary CFT, the number of these free
functions grows heavily with the total rank $r$ of the involved Jordan cells
and the number of logarithmic partner fields. However, there exist certain
permutation symmetries which relate many of these functions to each other.
The full derivation and further generalizations of our results will appear
elsewhere \cite{marco}.

\section{Ansatz for 4-point functions}
Let $r$ denote the rank of the Jordan cells we consider. One can show, that
in LCFTs with Jordan cells with respect to (at least) the $L_0$ mode, the
$h=0$ sector necessarily must carry such a Jordan cell structure. Furthermore,
its rank defines the maximal possible rank of all Jordan cells. Thus, without
loss of generality, we can assume that the rank of all Jordan cells is equal
to $r$, other cases can easily be obtained by setting certain structure
constants to zero. Each Jordan cell contains one proper highest weight state
giving rise to one proper irreducible subrepresentation. We will label this
state for a Jordan cell with conformal weight $h$ by $\ket{h;0}$. We choose
a basis in the Jordan cell with states $\ket{h;k}$, $k=0,\ldots,r-1$, such that
eq.~(\ref{eq:2}) is replaced by
\begin{equation}\label{eq:3}
        L_0\ket{h;k} = h\ket{h;k} + \ket{h;k-1}\ \ \textrm{for}\ \ 
        k=1,\ldots,r-1\,,\ \ \ \
        L_0\ket{h;0} = h\ket{h;0}\,.
\end{equation}
The corresponding fields will be denoted $\Psi_{(h;k)}$. Although the OPE
of two primary fields might produce logarithmic fields, we will further assume,
that primary fields {\slshape which are members of Jordan cells\/} are proper
primaries in the sense that OPEs among them only yield again primaries.

As discussed by Rohsiepe \cite{Rohsiepe:1996qj}, the possible structures
of indecomposable representations with respect to the Virasoro algebra
are surprisingly rich. Besides the defining condition eq.~(\ref{eq:3}),
further conditions have to be employed to fix the structure. The simplest
case is defined via the additional requirement
\begin{equation}\label{eq:L0qp}
       L_{1}\ket{h;k} = 0\,,\ \ \ \ 0\leq k<r\,.
\end{equation}
This condition means that all fields spanning the Jordan cell are
quasi-primary. It will be our starting point in the following. This condition
can be relaxed, but this will not concern us here.

Under these assumptions, as shown in \cite{Flohr:1997wm}, the action of the 
Virasoro modes receives an additional non-diagonal term.
The off-diagonal action is defined via
$\hat{\delta}_{h_i}\Psi_{(h_j;k_j)}(z) = \delta_{ij}\Psi_{(h_j;k_j-1)}(z)$ for
$k_j>0$ and $\hat{\delta}_{h_i}\Psi_{(h_j;0)}(z) = 0$. Thus,
\begin{equation}\label{eq:viir}
   L_n \eval{\Psi^{}_{(h_1;k_1)}(z_1)\ldots\Psi^{}_{(h_n;k_n)}(z_n)} =
   \sum_iz_i^n\left[z_i\partial_i + (n+1)(h_i+\hat{\delta}_{h_i})\right]
   \eval{\Psi^{}_{(h_1;k_1)}(z_1)\ldots\Psi^{}_{(h_n;k_n)}(z_n)}
\end{equation}
for $n\in\mathbb{Z}$.
Only the generators $L_{-1}$, $L_0$, and $L_1$ of the M\"obius
group admit globally valid conservation laws, which usually are expressed
in terms of the so-called conformal Ward identities
\begin{equation}\label{eq:ward} 0 = \left\{\begin{array}{rcl}
         L_{-1} G(z_1,\ldots z_n) & = & \sum_i\partial_i G(z_1,\ldots z_n)
         \,,\\[0.2cm]
         L_0 G(z_1,\ldots z_n) & = & \sum_i(z_i\partial_i + h_i +
         \hat{\delta}_{h_i})G(z_1,\ldots z_n)\,,\\[0.2cm]
         L_1 G(z_1,\ldots z_n) & = & \sum_i(z_i^2\partial_i
         + 2z_i[h_i + \hat{\delta}_{h_i}]) G(z_1,\ldots z_n)\,,
    \end{array}\right.
\end{equation}
where $G(z_1,\ldots z_n)$ denotes an arbitrary $n$-point function
$\eval{\Psi_{(h_1;k_1)}(z_1)\ldots\Psi_{(h_n;k_n)}(z_n)} \equiv
\eval{k_1k_2\ldots k_n}$ of primary
fields and/or their logarithmic partner fields. Here, we already have
written down the Ward identities in the form valid for proper
Jordan cells in logarithmic conformal field theories. Note that these are
now inhomogeneous equations. In principle,
we thus obtain a hierarchical scheme of solutions, starting with 
correlators of total Jordan-level $K=\sum_ik_i=r-1$, which fix
the generic form of all $n$-pt functions. In particular, correlators of 
solely proper primary fields vanish identically.

It is helpful to use a 
graphical representation where each field 
$\Psi_{(h;k)}(z)$ in a Jordan cell is depicted by a vertex with $k$ outgoing 
lines. Contractions of logarithmic fields give rise to logarithms in the
correlators, where the possible powers with which $\log(z_{ij})$ may occur
are determined by graph combinatorics.
$$\includegraphics[width=6cm]{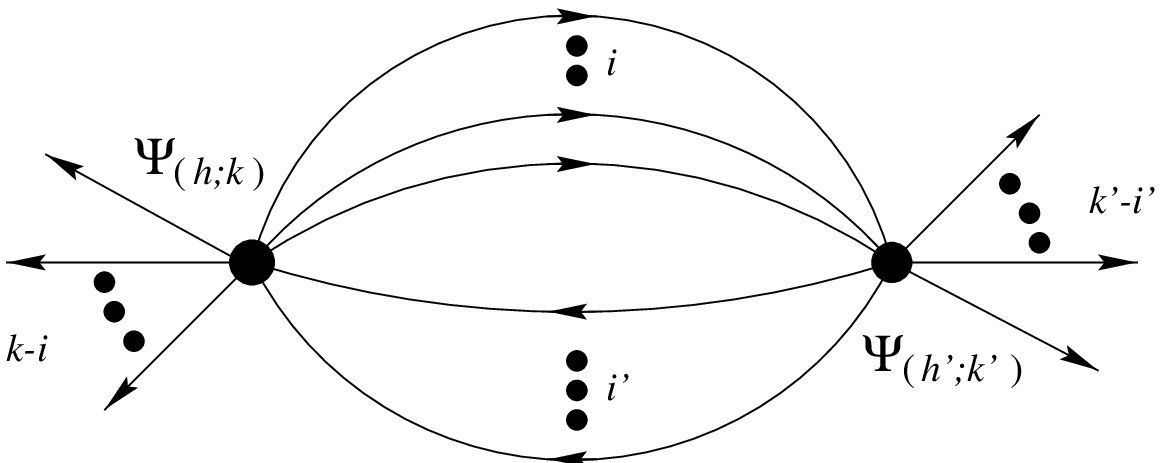}$$
Essentially, the terms of the generic of an $n$-pt function are given by a
sum over all admissible graphs subject to the rules\\
\mbox{$\bullet$\phantom{m}}Each vertex with $k_{{\rm out}}>0$ legs may at 
most receive $k'_{{\rm in}}i\leq r-1$ legs.\\
\mbox{$\bullet$\phantom{m}}Each vertex $i$ may only receive legs from vertices
$j\neq i$.\\
\mbox{$\bullet$\phantom{m}}A vertex for a proper primary field, $k_{{\rm out}}
=0$, does never receive legs.\\
\mbox{$\bullet$\phantom{m}}A total of exactly $r-1$ legs remains open, i.e.~
are not linked to other vertices.\\
Let us look at a small example. All admissible graphs, up to permutations, for
a 4-pt functions of a $r=2$ LCFT, where all fields are logarithmic, are given
by
$$\textrm{graphs}(\eval{1111})=\ 
  \raisebox{-0.6cm}{\includegraphics[width=6cm]{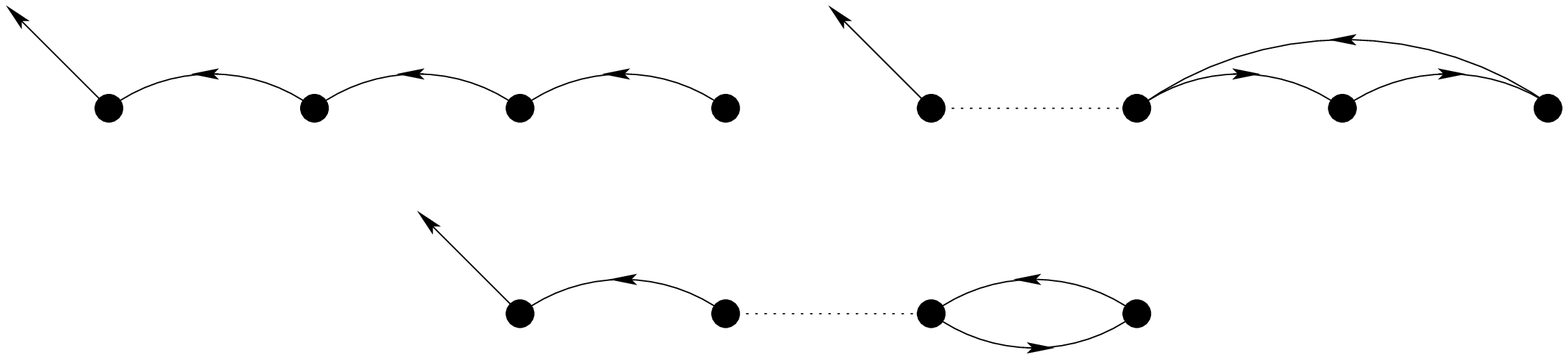}}\,.$$
Here, and in the following, we suppress all dependencies on the conformal
weights and coordinates and denote correlations functions simply by the set of
Jordan-levels $k_i$ of the fields $\Psi_{(h_i;k_i)}(z_i)$.

The linking numbers $A_{ij}(g)$ of a graph $g$ yield upper bounds on the 
powers, with which $\log(z_{ij})$ may occur. One needs a recursive
procedure to find all possible terms. One starts with the set of all different 
choices $f_i$ of $r-1$ legs which remain unlinked. Then, for all levels $K'$
and all choices $f_i$ one has to find all graphs which connect the remaining
$K-K'-(r-1)$ legs to vertices. Since the recurrence runs over $K'$, one can now
immediately write down the monomials in the $\log(z_{ij})$ as given by the
graphs $g$, multiplied with as yet undetermined constants $C(g)$. Imposing
global conformal invariance via eq.~(\ref{eq:ward}) fixes some of these 
constants. Further constraints stem from certain permutation symmetries, since
much of the generic structure of the correlators only depends on the
Jordan-levels, but not on which fields has what Jordan-levels.
Thus, the general ansatz is of the form
\begin{equation}
  \eval{k_1k_2k_3k_4} = \prod_{i<j}(z_{ij})^{\mu_{ij}}
        \sum_{(k'_1,k'_2,k'_3,k'_4)}\left[
        \sum_{g\in G_{K-K'}}
        C(g)
        \left(\prod_{i<j}\log^{A_{ij}(g)}(z_{ij})\right)\right]
        F_{k'_1k'_2k'_3k'_4}(x)\,,
\end{equation}
where $G_{K-K'}$ is set of graphs for $(k^{}_1-k'_1,\ldots,k^{}_4-k'_4)$,
$x=\frac{z_{12}z_{34}}{z_{14}z_{23}}$ is the crossing ratio, and
$\mu_{ij}=\frac{1}{3}\left(\sum_k h_k\right)-h_i-h_j$. 
Since the only explicit dependence on the conformal weights is through the
$\mu_{ij}$, one may put $h_1=\ldots=h_4=0$ for simplicity. A further convenient
abbreviation is $\ell_{ij}\equiv\log(z_{ij})$. Then, for $r=2$, one easily
finds
\begin{eqnarray*}
    \eval{1000} &=& F_{0}\,,\\ 
    \eval{1100} &=& F_{1100} - 2 \ell_{12} F_{0}\,,\\
    \eval{1110} &=& F_{1110} 
                      +(\ell_{12} -\ell_{13} - \ell_{23}) F_{1100} 
                      + (\ell_{13}-\ell_{12}-\ell_{23}) F_{1010}
                              - (\ell_{23}-\ell_{12}-\ell_{13}) F_{0110}\\
                & & {}+(-\ell_{12}^2-\ell_{13}^2-\ell_{23}^2 + 2 \ell_{12} 
                    \ell_{23} + 2 \ell_{12} \ell_{13} + 2 \ell_{23} \ell_{13}) 
                    F_{0}\\
                &=& F_{1110} + {\mathcal{P}_{(123)}}\left\{ 
                    (\ell_{12} - \ell_{23} - \ell_{13}) F_{1100} \right\}
                      +{\mathcal{P}_{(123)}} \left\{ \ell_{12} (\ell_{12} 
                    - \ell_{23} - \ell_{13})F_{0} \right\}\,.
\end{eqnarray*}
In the last case, we used the fact that the resulting form of the correlator
must obviously be invariant under permutation symmetry of the first three
Jordan-levels (not the first three fields!), which leads to identifications
between many of the $F_{k_1k_2k_3k_4}(x)$. The only caveat is that one has
to respect the ordering $i<j$ in all $z_{ij}$ or $\ell_{ij}$. Thus, in the
above example, $\mathcal{P}_{(123)}=(123)+(231)+(312)$ subject to the above 
rule. However, since
the full correlators, i.e.\ the correct single-valued combinations of
holomorphic and anti-holomorphic part, will only involve monomials in
$\log|z_{ij}|^2$, the ordering can be neglected for the $\ell_{ij}$.
In case that one considers the subset of correlators where all four
fields have the same conformal weight, $h_i=h_j$, one finds that
$$F_{k_1k_2k_3k_4}(x) = 
        F_{k_{\sigma(1)}k_{\sigma(2)}k_{\sigma(3)}k_{\sigma(4)}}(x)
        \ \ \ \ \forall\ \ \sigma\in S_4\,.$$
In the more general case with arbitrary conformal weights, $\sigma$ is
restricted to the subgroup of $S_4$, under which the original Jordan-levels on
the left hand side remain invariant. It is important to note that this still
implies identifications of structure functions with non-trivial exchanges of
their Jordan-level labels on the right hand side. Furthermore, we found cases
where an even higher symmetry can be implemented, identifying structure
functions with each other, which are not related at all by permutation
symmetries. We leave a full discussion to our forthcoming publication
\cite{marco}.

Unfortunately, even when all permutation symmetries are used to relate
different $F_{k_1k_2k_3k_4}$ with each other, explicit formul\ae\ easily
become very cumbersome, in particular for $r>2$. Alread the remaining case
for $r=2$, the correlator with four logarithmic fields, explodes to the
following monstrum, despite the fact that permutation symmetry has been
fully exploited:
  \begin{eqnarray*}
  \eval{1111} &=& F_{1111} +  {\mathcal{P}_{(1234)}} \big\{ \big[ 
                  (-\ell_{12} - \ell_{34} + \ell_{23} + \ell_{14}) {C_1}
                   +(\ell_{13} + \ell_{24} - \ell_{12} - \ell_{34}) {C_2}\\
              & & {\phantom{mmmmmmmmm}} 
                  - \ell_{14} + \ell_{34} - \ell_{13}) \big] F_{0111}\big\} \\
              &+& {\mathcal{P}_{(12)(34)}}
                  \big\{ \big[(\ell_{13}^2 +\ell_{24}^2 -
                  \ell_{14}^2 -\ell_{23}^2 + 2 (- \ell_{34} \ell_{24} - 
                  \ell_{12} \ell_{24} + \ell_{34} \ell_{14} + \ell_{13} 
                  \ell_{24} \\
              & & {\phantom{mr}}
                  - \ell_{13} \ell_{34}+ \ell_{23} \ell_{34}+ \ell_{12} 
                  \ell_{23} - \ell_{12} \ell_{13}- \ell_{23} \ell_{14}
                  + \ell_{12} \ell_{14})){C_3} \\
              & & {}+(-(\ell_{23} + \ell_{14})^2 +\ell_{23} \ell_{34}+\ell_{12} 
                  \ell_{14}-\ell_{13} \ell_{34}+\ell_{34} \ell_{14}+\ell_{13} 
                  \ell_{14} \\
              & & \phantom{mr}
                  -\ell_{34} \ell_{24} -\ell_{12} \ell_{13}-\ell_{12} 
                  \ell_{24}+\ell_{23} \ell_{24}+\ell_{23} \ell_{13}+\ell_{12} 
                  \ell_{23}+\ell_{24} \ell_{14})) {C_4}\\
              & & {}-\ell_{34}^2-\ell_{23}^2-\ell_{14}^2 +2 \ell_{23} 
                  \ell_{34} +2 \ell_{34} \ell_{14} -2 \ell_{12} \ell_{34} -
                  \ell_{23} \ell_{14}+\ell_{23} \ell_{24} \\
              & & \phantom{mr}
                  -\ell_{12} \ell_{13}+\ell_{12} \ell_{14}+\ell_{12} \ell_{23}-
                  \ell_{12} \ell_{24}+\ell_{13} \ell_{14}+\ell_{13} \ell_{24})
                  \big] F_{1100} \big\}\\
              &+& \big[ 2(\ell_{12} \ell_{24} \ell_{14} - \ell_{23} \ell_{13} 
                  \ell_{14}+ \ell_{23} \ell_{34} \ell_{24}- \ell_{24 } \ell_{13}
                  \ell_{34} - \ell_{23} \ell_{34} \ell_{14}  \\
              & & \phantom{mr}
                  -\ell_{12} \ell_{23} \ell_{34}- \ell_{12} \ell_{34} \ell_{24}
                  - \ell_{23} \ell_{13} \ell_{24}+ \ell_{12} \ell_{23} \ell_{13}
                  + \ell_{13} \ell_{34}\ell_{14} \\
              & & \phantom{mr}
                  -\ell_{13} \ell_{14} \ell_{24}- \ell_{23} \ell_{24} \ell_{14}
                  - \ell_{12} \ell_{13} \ell_{24}- \ell_{12} \ell_{23} \ell_{14}
                  - \ell_{12} \ell_{13} \ell_{34} 
                  -\ell_{12} \ell_{34} \ell_{14} ) \\
              & & {}+2( \ell_{13}^2 \ell_{24}+ \ell_{12}^2 \ell_{34}+ 
                  \ell_{14}^2 \ell_{23} + \ell_{23}^2 \ell_{14} + \ell_{34}^2 
                  \ell_{12}+ \ell_{24}^2 \ell_{13}) \big] F_0\,.
\end{eqnarray*}

\section{Permutation symmetries}
Note that certain constants still remain free. The reason for this becomes
apparent, when we use the graphical notation introduced earlier, which 
makes the structure of the formul\ae\ much more transparent and compact.
In essence, the polynomials in the $\ell_{ij}$ are completely symmetrized
with respect to their generating graphs. If we assume that all conformal
weights are equal, $h_i=h_j$, then the correlators enjoy even more
symmetries, since this means that it must be invariant under permutations
of the fields of the same Jordan-level. Due to limitation of space, we will
restrict ourselves here to this nice case and refer the reader to \cite{marco}
for the general case. Hence, we obtain, in our $r=2$ example,
\begin{eqnarray}
    \eval{1110} &= & 
      F_{1110} - {\mathcal{P}_{S_3}} (\ell_{12}) F_{0011} +  
      {\mathcal{P}_{S_3}} ( 2 \ell_{12} \ell_{23} -\ell_{12}^2 ) F_{0} 
      \nonumber\\
       &=& F_{1110} - {\mathcal{P}_{S_3}} ({\diagAone}\; ) F_{0011} +  
      {\mathcal{P}_{S_3}} ( 2 {\diagAtwo}\;  - {\diagCtwo}\, ) 
      F_{0} \,,\\
  \eval{1111} &= & F_{1111} - \tfrac{1}{6} {\mathcal{P}_{S_4}} 
  ({\diagAone}\;) F_{0111} + \tfrac{1}{4} {\mathcal{P}_{S_4}}  
  ({\diagAtwo}\; + {K_{S_4}^{(2)}} ) F_{0011} \nonumber\\
  & &{ }+ {\mathcal{P}_{S_4}} ( \tfrac{1}{2} {\diagEthree}\;+\tfrac{1}{3} 
    {\diagFthree}\; - {\diagCthree}\, ) F_0 \,.
\end{eqnarray}
Note that no free constants remained. On the other hand, to make the epxressions
as symmetric as possible, we encounter additional terms 
${K}\in\ker L_m^{\textrm{offdiag}}$ 
in the kernel of the nilpotent part of the Virasoro generators:
\begin{align}\ker(L_m-L_m')=&\
  \left\langle{K_1}\equiv\log(x),\
              {K_2}\equiv-\log(1-1/x)\right\rangle\nonumber\\
  =&\ \left\langle\ell_{12}+\ell_{34}-\ell_{14}-\ell_{23},\
               \ell_{12}+\ell_{34}-\ell_{13}-\ell_{24}\right\rangle\,,\\
   {K^{(2)}_{S_4}}=&\ {K_1^2 -K_1K_2+K_2^2}\,.
\end{align}
Here, $L_m'$ denotes the ordinary part of the Virasoro mode without the
off-diagonal action $\hat{\delta}$ from eq.~(\ref{eq:viir}).
In principle, all 4-pt functions for arbitrary rank $r$ LCFTs can be
computed in this way. We conclude these notes with a few examples for
rank $r=3$ and, for the sake of simplicity, all conformal weights identical.
Each individual Jordan level $k_i$ may now vary in the range $0\leq k_i\leq 2$.
We find
\begin{eqnarray}
    \eval{2000}\ \ =\ \ \eval{1100} &=& F_0\,,\\
    \eval{2100} &=& F_{2100} - 2\ell_{12}F_0\,,\\
    \eval{1110} &=& F_{1110} - (\ell_{12} + \ell_{23} + \ell_{13}) F_{0} \\
                &=& F_{1110} - {\mathcal{P}_{(123)}}\left\{\ell_{12} 
                    F_{0} \right\}\,,
\end{eqnarray}
where the first line is due to the general result that the lowest total
Jordan level with non-vanishing correlator is $K=r-1$ and that this
correlator looks like a correlator of four primary fields in ordinray
non-logarithmic CFT. A more involved correlator for $r=3$ is the following,
where again a kernel term shows up:
\begin{equation}
    \eval{1111} =  F_{1111} -\frac 16 
    {\mathcal{P}_{S_{4}}} ( \ell_{12} )  F_{0111} + 
       \Big\{ {\mathcal{P}_{S_{4}}} \Big[ -\frac 14 \ell_{34}^2 + 
    \frac 12 \ell_{34} \ell_{24} \Big] + {K^{(2)}_{S_4}} \Big\} F_0
    \,.
\end{equation}
Note the similarity to the $r=2$ case. With $K_\pm = K_1\pm K_2$ and
${\mathcal{P}_4} = {1 + P_{12} + P_{34} + P_{12} P_{34}}$, a really
non-trivial example, where a non-graphical expansion would fill several
pages, is
\begin{align}
  \phantom{m}\eval{2211} = &\ F_{2211} \nonumber\\[-1mm]
 +&\ {\mathcal{P}_4} \Big\{ -\tfrac 12 \perm{(13)(24)} {\diagAone}\; 
   + {K_+}\Big\} F_{0122} + 
   {\mathcal{P}_4} \Big\{ -\tfrac 12 {\diagAone}\; 
   + {K_+}\Big\} F_{1112} \nonumber\\[-1mm]
 +&\ {\mathcal{P}_4} \Big\{  \left[\tfrac{1}{2} \perm{(24)} 
   -\tfrac{1}{6} \right]
   {\diagAtwo}\;+ \left[ \tfrac{1}{3}  -\tfrac{1}{2} \perm{(14)} \right] 
   {\diagBtwo}\;+
   \tfrac{1}{12}  {\diagCtwo}\; + {K^{(2)}_{S_4}} \Big\} 
   F_{1111} \nonumber\\[-1mm]
 +&\ {\mathcal{P}_4} \Big\{ \left[ \tfrac{1}{2} -\tfrac{1}{2} \perm{(24)} 
   \right] 
   {\diagAtwo}\;+ \left[ \perm{(12)}+\tfrac{1}{2} \perm{(14)} \right] 
   {\diagBtwo}\; - 
   \tfrac{1}{4}  {\diagCtwo}\; + {K^{(2)}_{S_4}} + 
   {K_1 K_2} \Big\} 
   F_{0112} \nonumber \\[-1mm]
 +&\ {\mathcal{P}_4} \Big\{ \left[\tfrac{1}{2} \perm{(23)} -\tfrac{1}{2}  
   \right] 
   {\diagAtwo}\; +\left[\perm{(24)}-\tfrac{1}{2} \perm{(243)} \right]
   {\diagBtwo}\; - 
   \tfrac{1}{4} \perm{(13)(24)}{\diagCtwo}\; + {K_{-}^2}    
   \Big\} F_{0022} \nonumber\\[-1mm]
 +&\ {\mathcal{P}_4} \Big\{  \left[  -\tfrac{1}{2} - 2 \perm{(14)} \right] 
   {\diagEthree}\; + \left[ 2\perm{(243)}+\perm{(24)}-\perm{(13)} \right] 
   {\diagCthree} \nonumber \\[-1mm]
  &\ \qquad {}-\perm{(14)} {\diagFthree}\; + 2 {\diagAthree}\; - 
   {\diagBthree}  - 
   \tfrac{1}{2}  {\diagDthree}\; + 
   {K_{-}^2 K_{+}}  \Big\} F_{0012} \nonumber\\[-1mm]
 +&\ {\mathcal{P}_4} \Big\{ 
   \left[\tfrac 16 \perm{134} - \tfrac 76 - \tfrac 16 
   \perm{(13)}  \right] {\diagAthree}\; - \left[ 2 \perm{(23)} + \tfrac 56 
   \perm{(234)} + \tfrac{1}{2} \perm{(132)} \right] 
   {\diagCthree} \nonumber \\[-1mm]
  &\ \qquad {}+\left[ \tfrac{1}{12} + \tfrac{11}{6} \perm{(23)} \right] 
   {\diagEthree}\; + \left[ \tfrac 16 + \tfrac 56 \perm{(14)} \right] 
   {\diagFthree}\; + 
   \tfrac 12 {\diagBthree}  +\tfrac 13 {\diagDthree}\; 
   + {K_{-}^2 K_{+}} 
   \Big\} F_{0111} \nonumber \\[-1mm]
 +&\ {\mathcal{P}_4} \Big\{  
   \left[ \tfrac 12 \perm{(12)(34)} \!+\! \tfrac 12 
   \perm{(1243)} \!-\! 2 \perm{(124)} \!-\! \tfrac 12 \perm{(142)} \!-\! 2 
   \perm{(143)} \right] {\diagDfour}\; + \tfrac{3}{16} 
   {\diagEfour} \nonumber\\[-1mm]
  &\ \qquad {}+\left[\tfrac 12 \perm{(34)}  \!+\!  \tfrac 12 \perm{(123)} 
   \!-\! \tfrac 12 \perm{(134)} \right]  {\diagAfour}\; + \left[ \tfrac 34 
   \perm{(234)} \!-\! \perm{(1243)} \!+\! \tfrac 54 \perm{(132)} \right] 
   {\diagBfour} \nonumber\\[-1mm]
  &\ \qquad {}+\left[ \tfrac 14 \perm{(12)(34)} \!-\! \tfrac 12 \perm{(23)} 
   \!+\! \tfrac 12 \perm{(1234)} \!-\! \tfrac 18 \perm{(14)}  \right] 
   {\diagFfour}\; + \left[ \tfrac 58 \perm{(24)} \!-\! \tfrac 18 \right] 
   {\diagIfour} \nonumber\\[-1mm]
  &\ \qquad {}+\left[ \perm{(234)} \!-\! \tfrac 12 \perm{(243)} \!-\! \tfrac 34 
   \perm{(14)} \right] {\diagCfour}\;  + \left[ 2 \perm{(23)} \!+\! 
   \perm{(14)} \!-\! \perm{(1324)} \right] {\diagJfour} \nonumber\\[-1mm]
  &\ \qquad {}+\left[ \tfrac 12 \perm{(123)} \!-\! \perm{(12)(34)} \!-\! 
   \tfrac 12 \perm{(1234)} \right] {\diagGfour}\; 
   + \tfrac 12 \perm{(23)} {\diagHfour}\; + 
   {\diagKfour}\; +
   {K_{-}^2 K_1 K_2} \Big\} F_{0}\,. 
\end{align} 
To summarize, the computational complexity grows heavily with
the rank $r$ and total Jordan level $K$. Already the
generic solution for $r=2$ and $K_{\textrm{max}}=
4(r-1)=4$ needs a computer program. The form of 4-pt functions, 
as determined by global
conformal invariance, is much more complicated than in the ordinary case and
crossing symmetry must explicitly be taken into account to fix it.
There exist additional degrees of freedom ${\ker(L_m-L_m')}$
not present in ordinary CFT.

{\bf Acknowledgement} The research of M.\,F.\ is supported by
the European Union network HPRN-CT-2002-00325 and the research of M.\,F.\ and
M.\,K.\ is supported by the string theory network
(SPP no.\ 1096), Fl 259/2-2, of the Deutsche Forschungsgemeinschaft.

\end{document}